# Correct use of the Gordon decomposition in the calculation of nucleon magnetic dipole moments

Mustapha Mekhfi

International Center for Theoretical Physics, Trieste, Italy

And

Département de Physique, Université Es-senia, Algerie

We perform the calculation of the nucleon dipole magnetic moment in full details using the Gordon decomposition of the free quark current. This calculation has become necessary because of frequent misuse of the Gordon decomposition by some authors in computing the nucleon dipole magnetic moment.

The dipole magnetic moment operator of the nucleon is defined as follows:

$$\vec{\mu}_N = \sum_{i=u,d,s} \int \frac{Q_i}{2} d^3x \, \bar{\psi}_i \vec{x} \times \vec{\gamma} \psi_i + \cdots \quad (1.1)$$

Ellipses indicate contributions to the dipole magnetic moment from other sources than the spin of the quarks. In the Chromodynamics framework, the contribution could be due to quark anti quarks pairs to which the probing photon couples. Also, in chiral models framework, the contribution could be due to charged mesons to which the probing photon couples. These contributions are generally small and are of no concern here. Therefore, their study is postponed to a forthcoming work. In case only quark creation and annihilation operators are considered in intermediary steps, the quark field as follows:

$$\psi_i(x) = \sum_s \int \frac{d^3k}{(2\pi)^3} \sqrt{\frac{m_i}{k_0}} a_{i,k,s} u_{i,k,s} \exp(-i\vec{k}.\vec{x})$$

$$u_{i,k,s} = \begin{pmatrix} \sqrt{\frac{k_0 + m_i}{2m_i}} \chi \\ \sqrt{\frac{k_0 - m_i}{2m_i}} \frac{\vec{k}}{k} . \vec{\sigma} \chi \end{pmatrix} \quad (1.2)$$

Thus, the magnetic moment takes the form:

$$i \sum_s \int \vec{\nabla}_q \times (\bar{\psi}_{i,s}(k') \vec{\gamma} \psi_{i,s}(k)) \frac{d^3K}{(2\pi)^3} |_{\vec{q}=0}$$

$$\vec{K} = \frac{\vec{k} + \vec{k}'}{2}, \quad \vec{q} = k - k' \quad (1.3)$$

In what follows, we will omit flavor and spin indices to simplify notations, we shall write $\psi(k) = \sqrt{\frac{m}{k_0}} u_k$. Equation(1.3) is the starting formula used before applying Gordon decomposition. Note that one has to first perform the differentiation with respect to $\vec{q}$ before setting $\vec{q} = 0$. The Gordon decomposition of the quark current is given by the well known formula:

$$\bar{\psi}(k')\gamma^\mu \psi(k) = (\bar{\psi}(k')\left[\frac{(k'+k)^\mu}{2m} + i\sigma^{\mu\nu}\frac{(k'-k)_\nu}{2m}\right]\psi(k) \qquad (1.4)$$

We start by inserting the first term of the decomposition (convection term) into the magnetic moment formula (1.3):

$$i\int \left[\vec{\nabla}_q ((\bar{\psi}(k')\psi(k)) \times \frac{\vec{K}}{m}\right]_{\vec{q}=0} \frac{d^3K}{(2\pi)^3}$$

The gradient of the above product has to be developed by taking into account that the dependencies on $k$ and $k'$ are both present in the creation, annihilation operators $a_{k'}^\dagger, a_k$ as well as in the Dirac spinors $u^\dagger$, $u$:

$$\vec{\nabla}_q (a_{k'}^\dagger a_k \bar{u}_{k'} u_k) = \vec{\nabla}_q (a_{k'}^\dagger a_k) \bar{u}_{k'} u_k + a_{k'}^\dagger a_k \vec{\nabla}_q (\bar{u}_{k'} u_k)$$

In the limit $\vec{q} = 0$, the above equation reads:

$$\vec{\nabla}_q (a_{k'}^\dagger a_k \bar{u}_{k'} u_k)\big|_{\vec{q}=0} = \vec{\nabla}_q (a_{k'}^\dagger a_k)\big|_{\vec{q}=0} + a_{k'}^\dagger a_k \vec{\nabla}_q (\bar{u}_{k'} u_k)\big|_{\vec{q}=0} \qquad (1.5)$$

The first term in the above expression relates to the non relativistic quark orbital angular momentum which is put zero since we consider that the nucleon is in the ground state. In computing this term, we may use the standard relations $\bar{u}_k u_k = 1$ and $a_{k'}^\dagger a_k + a_k a_{k'}^\dagger = \delta_{\vec{k},\vec{k}'}$. Due to the small component of the Dirac wave function $\sim \frac{\vec{k}.\vec{\sigma}}{k}$, the second term in (1.5) is non-vanishing and represents the relativistic contribution to the convection term. In computing this term, we write first $\vec{\nabla}_q = \frac{1}{2}(\vec{\nabla}_k - \vec{\nabla}_{k'})$, $d^3K\big|_{\vec{q}=0} = d^3k$. After integrating by part, we get:

$$\int \vec{\nabla}_q (\bar{u}_k \cdot u_k)\big|_{\vec{q}=0} \times \vec{k} \frac{d^3k}{(2\pi)^3} = \int \bar{u}_k \vec{\nabla}_k (u_k) \times \vec{k} \frac{d^3k}{(2\pi)^3} \qquad (1.6)$$

We then calculate the gradient of the spinor $u$ using its explicit expression in (1.2):

$$\vec{\nabla}_k (u_k) = -\frac{1}{k_0 + m} \vec{\gamma} u_{kL} + ...$$

$$u_{kL} = \frac{(1+\gamma_0)}{2} u_k = \begin{pmatrix} \sqrt{\frac{k_0+m}{2m}} \chi \\ 0 \end{pmatrix}$$

Where ellipses indicate terms proportional to $\vec{k}$ and $u_{kL}$ designates the large component of the spinor. The terms that are proportional to $\vec{k}$ give vanishing contributions in (1.6) since $\vec{k} \times \vec{k} = 0$. We insert this result into (1.6), approximate the factor $k_0$ by the average value of relativistic quark energy inside the nucleon, and get (we keep the notation $k_0$ to designate the average value):

$$-\frac{i}{k_0 + m} \int \bar{\psi}_k \vec{k} \times \vec{\gamma} \psi_L \frac{d^3k}{(2\pi)^3} \qquad (1.7)$$

At this stage of the computation we make use of the relation:

$$\vec{\gamma} = \vec{n}(\vec{n}.\vec{\gamma}) - i\,(\vec{n}\vec{\gamma})(\vec{n} \times \vec{\gamma})\gamma_0 \gamma_5$$

$$\text{where } \vec{n} = \frac{\vec{k}}{|\vec{k}|} \qquad (1.8)$$

This formula is the generalization of the well known relation of Pauli matrices $\vec{\sigma} = (\vec{n}.\vec{\sigma})\vec{n} - i(\vec{n}\vec{\sigma})(\vec{n} \times \vec{\sigma})$. We insert (1.8) into (1.7) and define $x = \frac{m}{k_0}$. After some algebra, we get:

$$\begin{aligned}
-\frac{i}{k_0+m}\int \bar{\psi}_k \vec{k}\times\vec{\gamma}\psi_L \frac{d^3k}{(2\pi)^3} &= \frac{x-1}{x+1}\int \bar{\psi}\vec{\gamma}\gamma_5 \psi_L \frac{d^3k}{(2\pi)^3} - \int \frac{\bar{\psi}\gamma_5 \psi_L \vec{k}}{k_0+m}\frac{d^3k}{(2\pi)^3} \\
&= \frac{1}{2}\frac{x-1}{x+1}\int (\bar{\psi}\vec{\gamma}\gamma_5\psi + \bar{\psi}\vec{\gamma}\gamma_5\gamma_0\psi)\frac{d^3k}{(2\pi)^3} - \int \frac{\bar{\psi}\gamma_5\psi_L \vec{k}}{k_0+m}\frac{d^3k}{(2\pi)^3} \\
&= \frac{x-1}{x+1}(\vec{S}+\frac{\vec{\delta}}{2}) - \int \frac{\bar{\psi}\gamma_5\psi_L \vec{k}}{k_0+m}\frac{d^3k}{(2\pi)^3}
\end{aligned} \qquad (1.9)$$

The first and the second term in brackets are interpreted respectively as the transverse and the longitudinal spin:

$$\vec{S} = \frac{1}{2}\int \bar{\psi}\vec{\gamma}\gamma_5\psi \frac{d^3k}{(2\pi)^3}$$

$$\vec{\delta} = \int \bar{\psi}\vec{\gamma}\gamma_5\gamma_0\psi \frac{d^3k}{(2\pi)^3}$$

The remaining term $\bar{\psi}\gamma_5\psi_L$ has to be worked out further. To this end, we write:

$$-\int \frac{\bar{\psi}\gamma_5\psi_L \vec{k}}{k_0+m}\frac{d^3k}{(2\pi)^3} = \int \frac{\chi^\dagger(\vec{\sigma}.\vec{k})\vec{k}\chi}{2k_0(k_0+m)}\frac{d^3k}{(2\pi)^3}$$

Then, we decompose the scalar $\chi^\dagger(\vec{\sigma}.\vec{k})\chi$ in terms of longitudinal and transverse spins. At this point, we reestablish flavor indices to indicate that field operators act at the quark level. Let us, firstly, make explicit the spin expressions:

$$\vec{S}_i = \frac{1}{2}\sum_{s's}\int d^3k\, \chi^\dagger(\vec{\sigma} - \frac{i\vec{\sigma}.\vec{k}\vec{\sigma}\times\vec{k}}{k_0(m_i+k_0)})\chi a^\dagger_{i,k,s'}a_{i,k,s} + \cdots$$

$$\vec{\delta}_i = \sum_{s's}\int d^3k\, \chi^\dagger(\frac{m_i}{k_0}\vec{\sigma} + \frac{\vec{\sigma}.\vec{k}}{k_0(m_i+k_0)}i\vec{\sigma}\times\vec{k})\chi a^\dagger_{i,k,s'}a_{i,k,s} + \cdots \quad (1.10)$$

$$= \sum_{s's}\int d^3k\, \chi^\dagger(\vec{\sigma} - \frac{(\vec{\sigma}.\vec{k})\vec{k}}{k_0(m_i+k_0)})\chi a^\dagger_{i,k,s'}a_{i,k,s} + \cdots$$

Ellipses in the following expressions denote anti quarks which we omit for simplicity, and pair creation (annihilation) terms which we neglect. We will use the second expression of the transverse spin above. However, the longitudinal spin will be rearranged using the following relation among Pauli matrices:

$$i\vec{\sigma}.\vec{k}\vec{\sigma}\times\vec{k} = (k_0^2 - m^2)\vec{\sigma} - \vec{\sigma}.\vec{k}\,\vec{k}$$

This leads to a new expression for the longitudinal spin (of interest to us). (1.10) becomes:

$$\vec{S}_i = \frac{1}{2}\sum_{s's}\int d^3k\,\chi^\dagger(x_i\vec{\sigma}+\frac{\vec{\sigma}.\vec{k}\,\vec{k}}{k_0(m_i+k_0)})\chi a^\dagger_{i,k,s'}a_{i,k,s}+\cdots$$

$$\vec{\delta}_i = \sum_{s's}\int d^3k\,\chi^\dagger(\vec{\sigma}-\frac{(\vec{\sigma}.\vec{k})\vec{k}}{k_0(m_i+k_0)})\chi a^\dagger_{i,k,s'}a_{i,k,s}+\cdots$$

Multiplying the transverse spin by the factor $-\frac{x_i}{2}$ and then adding the resulting expression to the longitudinal spin expression, we extract the term of interest $\sim \vec{\sigma}.\vec{k}\,\vec{k}$.

$$\sum_{s's}\int d^3k\,\chi^\dagger(\frac{(\vec{\sigma}.\vec{k})\vec{k}}{2k_0(m_i+k_0)})\chi a^\dagger_{i,k,s'}a_{i,k,s} = \frac{1}{(1+x_i)}(\vec{S}_i - \frac{x}{2}\vec{\delta}_i) \qquad (1.11)$$

Inserting (1.11) into (1.9) we get:

$$\frac{x_i-1}{x_i+1}(\vec{S}_i+\frac{\vec{\delta}_i}{2})+\frac{1}{(1+x_i)}(\vec{S}_i-\frac{x_i}{2}\vec{\delta}_i) = \frac{x_i}{x_i+1}(\vec{S}_i-\frac{\vec{\delta}_i}{2x_i})$$

We define the nucleon dipole magnetic moment $\vec{\mu}_N$, the longitudinal spin distribution $\Delta_q$ and the transverse spin distribution $\vec{\delta}$ respectively as follows:

$$\vec{\mu}_N = \langle PS|\sum_{i=u,d,s}\int\frac{Q_i}{2}dx^3\bar{\psi}_i\vec{x}\times\vec{\gamma}\psi_i|PS\rangle+\cdots$$

$$\Delta_i\vec{S} = \langle PS|\frac{1}{2}\int dx^3\bar{\psi}_i\vec{\gamma}\gamma_s\psi_i|PS\rangle$$

$$\vec{\delta}_i = \langle PS|\int dx^3\psi^\dagger_i\vec{\gamma}\gamma_s\psi_i|PS\rangle$$

For the transverse spin, the same notation is currently used to express an operator acting at the quark and at the nucleon level. The relatively long computation of the "convection current" contribution to the nucleon dipole magnetic moment[1,2] comes to an end. For a nucleon at rest and polarized along the third direction, this contribution reads:

$$\mu_N = \sum_{i=u,d,s}\frac{1}{2}\frac{x_i\mu_i}{x_i+1}\langle P\uparrow|2(\vec{S}_i)_3-\frac{(\vec{\delta}_i)_3}{x_i}|P\uparrow\rangle+\cdots$$

$$= \sum_{i=u,d,s}\frac{1}{2}\frac{x_i\mu_i}{x_i+1}(\Delta_i-\frac{\delta_i}{x_i})+\cdots$$

$$\mu_i = \frac{Q_i}{2m_i}$$

Where $\Delta_i$ are the quark contribution to the baryon spin and $\delta_i$ the baryon tensor charge.

It remains to work out the spin part. For this, we return back to the Gordon decomposition in (1.4) and insert the spin part into (1.3):

$$-\frac{1}{2m}\int\left[\vec{\nabla}_q\times((\bar{\psi}(k')q_\nu\vec{\sigma}^\nu\psi(k))\right]_{\vec{q}=0}\frac{d^3k}{(2\pi)^3} \quad (1.12)$$

Where $\vec{\sigma}^\nu$ is a vector matrix which components are $\sigma^{i\nu}, i=1,2,3$. Then, we write:

$$\begin{aligned}-\vec{\sigma}^\nu q_\nu &= -\vec{q}\times\vec{\Sigma}+i\vec{\alpha}q_0 \\ \sigma^{ij} &= \epsilon^{ijk}\Sigma_k \\ &= \epsilon^{ijk}\gamma_k\gamma_5\gamma_0 \\ \sigma^{i0} &= -i\alpha^i\end{aligned} \quad (1.13)$$

We differentiate the term $-\vec{q}\times\vec{\Sigma}$ and obtain

$$\begin{aligned}\vec{\nabla}_q\times(\vec{q}\times\bar{\psi}\vec{\Sigma}\psi)\big|_{\vec{q}=0} &= -2\bar{\psi}\vec{\Sigma}\psi \\ &= -2\bar{\psi}\vec{\gamma}\gamma_5\gamma_0\psi\end{aligned} \quad (1.14)$$

Where the following identity has been used:

$$\begin{aligned}\vec{\nabla}_q\times(\vec{q}\times\bar{\psi}\vec{\Sigma}\psi) = \\ \vec{q}\vec{\nabla}_q.(\bar{\psi}\vec{\Sigma}\psi)-(\vec{q}.\vec{\nabla}_q)(\bar{\psi}\vec{\Sigma}\psi)-\bar{\psi}\vec{\Sigma}\psi\vec{\nabla}_q.\vec{q}+(\bar{\psi}\vec{\Sigma}\psi.\vec{\nabla}_q)\vec{q}\end{aligned}$$

The first and the second term will vanish, when setting $\vec{q}$ to zero.

We differentiate now the term $i\vec{\alpha}q_0$ of (1.13). This term is subject to frequent misuses by some authors [3] [4] [5] who use some arguments to drop it. This term has motivated our present need to re-compute the quark dipole magnetic moment of the nucleon more carefully:

$$\left[i\nabla_{\vec{q}}\times(\bar{\psi}\vec{\alpha}q_0\psi)\right]_{\vec{q}=0}$$

Effectively, the assumption made by these authors amounts to placing wrongly the $q_0$ outside the gradient $\vec{\nabla}_q$:

$$\left[ iq_0 \vec{\nabla}_{\vec{q}} \times (\bar{\psi} \vec{\alpha} \psi) \right]_{\vec{q}=0} = 0$$

Written as above, it is obviously vanishing since $q_0 = \sqrt{k^2 + m^2} - \sqrt{k'^2 + m^2}$ and $\vec{q} = 0$ implies $k = k'$. The correct expression of this time derivative term should be with $q_0$ inside the gradient $\vec{\nabla}_{\vec{q}}$. To compute this litigious term we use the $\gamma$ matrices identity:

$$\frac{\vec{\gamma}(\vec{\gamma}.\vec{k}) - (\vec{\gamma}.\vec{k})\vec{\gamma}}{2} = i(\vec{\gamma} \times \vec{k}) \gamma_0 \gamma_5$$

And get successively:

$$\left[ i\vec{\nabla}_{\vec{q}} \times (\bar{\psi}\vec{\alpha}q_0\psi) \right]_{\vec{q}=0} = i\frac{\vec{k} \times \bar{\psi}\vec{\alpha}\psi}{k^0} \quad (1.15)$$
$$= x\bar{\psi}\vec{\gamma}\gamma_5\psi - \bar{\psi}\vec{\gamma}\gamma_5\gamma_0\psi$$

The second line in (1.15) is obtained by using the Dirac equation for both $\psi$ and $\bar{\psi}$.

Inserting both (1.15) and (1.14) into (1.13) we get.

$$2\bar{\psi}\vec{\gamma}\gamma_5\gamma_0\psi + x\bar{\psi}\vec{\gamma}\gamma_5\psi - \bar{\psi}\vec{\gamma}\gamma_5\gamma_0\psi = x(\bar{\psi}\vec{\gamma}\gamma_5\psi + \frac{\bar{\psi}\vec{\gamma}\gamma_5\gamma_0\psi}{x}) \quad (1.16)$$

Then inserting the result into (1.12), we get the spin part contribution to the nucleon dipole magnetic moment:

$$\mu_N = \sum_{i=u,d,s} \frac{\mu_i x_i}{2} \left\langle P\uparrow \left| 2(\vec{S}_i)_3 + \frac{(\vec{\delta}_i)_3}{x_i} \right| P\uparrow \right\rangle + \cdots$$

It is worth to note that the convection and the spin part are associated respectively to the orthogonal combinations $(\Delta_i - \frac{\delta_i}{x})$ and $(\Delta_i + \frac{\delta_i}{x})$. Finally, by adding anti quarks contribution to the nucleon dipole magnetic moment, we get an expression in terms of two measurable quantities describing the spin in the relativistic regime, namely $\Delta_{q,\bar{q}}$, and $\delta_{q,\bar{q}}$.

$$\mu_N = \sum_{i=u,d,s} \frac{x_i \mu_i}{2} \left( \frac{1}{x_i+1}(\Delta_i - \Delta_{\bar{i}} - \frac{\delta_i - \delta_{\bar{i}}}{x_i}) + (\Delta_i - \Delta_{\bar{i}} + \frac{\delta_i - \delta_{\bar{i}}}{x_i}) \right) + \cdots \quad (1.17)$$

Where the first term in the above expression comes from the convection current and the second term comes from the spin part of the Gordon decomposition of the quark current. We checked that our derivation is correct by rearranging our terms in equation (1.17) and find an expression that has been already derived by other authors (see for instance reference[6]):

$$\mu_N = \sum_{i=u,d,s} x_i \mu_i \left[ (1 - \frac{x_i}{2(1+x_i)})(\Delta_i - \Delta_{\bar{i}}) + \frac{1}{1+x_i} \frac{1}{2}(\delta_i - \delta_{\bar{i}}) \right] + \cdots (1.18)$$

[1] M Mekhfi, Phys Rev D **72**, 114014 (2005).

[2] M Mekhfi, 17th International Spin Physics Symposium, Kyoto, Japan, , Vol 915,P 642, (2006), edited by AIP Proceeding, High Energy Physics Springer, (2006).

[3] X Artru, Proceedings of RIKEN BNL( Workshop on Future Measurements), Vol 29, (2000), edited by D.Boer and M.G.Perdekamp

[4] X.S Chen, D Qing, W.M Sun, H.S Zong, and F Wang, Phys Rev C **69**, 045201 (2004).

[5] L-J Dai, W-M Sun, D Qing, H-S Zong, and X-S Chen, Phys Rev D **74**, 017901 (2006).

[6] Di Qing, Xiang-Song Chen, and Fan Wong, Phys.Rev.D **58**, 114032 (1998).